# Energy spectrum of the relativistic Dirac-Mörse problem

**A. D. Alhaidari**

Physics Department, King Fahd University of Petroleum & Minerals, Box 5047,
Dhahran 31261, Saudi Arabia
E-mail: haidari@mailaps.org

**Abstract:** We derive an elegant analytic formula for the energy spectrum of the relativistic Dirac-Mörse problem which has been solved recently. The new formula displays the properties of the spectrum more vividly.

Exact solutions of Dirac equation with potential interaction are very rare. Since the original work of Dirac in the early part of last century up until 1989 only the relativistic Coulomb problem was solved exactly. In 1989, the relativistic extension of the oscillator problem (Dirac-Oscillator) was finally formulated and solved exactly by Moshinsky and Szczepaniak [1]. Recently, the Dirac-Mörse problem was also solved [2], its bound states spectrum and spinor wavefunctions were obtained. The energy spectrum was written as a solution to a quadratic equation involving the physical parameters of the problem [equation (6) below]. In this letter, we reproduce this spectrum and write it in an alternative but equivalent form. The new analytic formula [equation (5) below] is distinctly elegant and gives a better picture of the overall behavior of the spectrum.

The relativistic potential in Dirac equation for problems with spherical symmetry has in general two components, an "even" and an "odd" component. They are related by a "gauge fixing condition" [2,3]. The wave equation gives two coupled first order differential equations for the two radial spinor components. By eliminating one component we obtain a second order differential equation for the other. The resulting equation may turn out not to be Schrödinger-like, i.e. it may contain first order derivatives. We apply a global unitary transformation to eliminate the first order derivatives. This produces a constraint that relates the even and odd components of the relativistic potential (the "gauge fixing condition"). Comparison of the resulting Schrödinger-like equation with well-known exactly solvable nonrelativistic problems gives a correspondence map among the potential parameters, angular momentum, and energy of the relativistic and nonrelativistic problem. Using this parameter substitution map and the energy spectrum of the nonrelativistic problem one can easily and directly obtain the relativistic spectrum. By employing this strategy, it was also possible to obtain new exact solutions of Dirac equation that are the relativistic extension of yet another class of shape invariant potentials [3], which includes Dirac-Pöschl-Teller, Dirac-Eckart, etc.

Using gauge invariance in QED, it was shown [2,3] that the radial Dirac equation for a charged spinor in spherically symmetric electromagnetic field could generally be written as

$$\begin{pmatrix} 1+\alpha^2\mathcal{V}(r) & \alpha\left(\dfrac{\kappa}{r}+\mathcal{W}(r)-\dfrac{d}{dr}\right) \\ \alpha\left(\dfrac{\kappa}{r}+\mathcal{W}(r)+\dfrac{d}{dr}\right) & -1+\alpha^2\mathcal{V}(r) \end{pmatrix}\begin{pmatrix}\phi(r)\\ \theta(r)\end{pmatrix}=\varepsilon\begin{pmatrix}\phi(r)\\ \theta(r)\end{pmatrix} \qquad (1)$$

where $\alpha$ is the fine structure constant, $\varepsilon$ is the relativistic energy and $\kappa$ is the spin-orbit coupling parameter defined as $\kappa = \pm (j + \frac{1}{2})$ for $l = j \pm \frac{1}{2}$. The real radial functions $\mathcal{V}(r)$ and $\mathcal{W}(r)$ are the even and odd components of the relativistic potential, respectively. $\mathcal{W}(r)$ is a gauge potential that does not contribute to the magnetic field. However, fixing this gauge degree of freedom by taking $\mathcal{W} = 0$ is not the best choice. An alternative and proper "gauge fixing condition", which is much more fruitful, will be imposed as a constraint in equation (2) below. Now, equation (1) gives two coupled first order differential equations for the two radial spinor components. By eliminating the lower component we obtain a second order differential equation for the upper. To make this equation Schrödinger-like, we apply a global unitary transformation parameterized by a real angle $\rho$ that eliminates the first order derivatives [2,3]. This results in a constraint that relates the even and odd components of the relativistic potential as follows:

$$\mathcal{V}(r) = \frac{S}{\alpha}\left[\mathcal{W}(r) + \frac{\kappa}{r}\right] \qquad (2)$$

It also maps the radial Dirac equation (1) into

$$\begin{pmatrix} C+2\alpha^2\mathcal{V} & \alpha\left(-\dfrac{S}{\alpha}+\dfrac{\alpha C}{S}\mathcal{V}-\dfrac{d}{dr}\right) \\ \alpha\left(-\dfrac{S}{\alpha}+\dfrac{\alpha C}{S}\mathcal{V}+\dfrac{d}{dr}\right) & -C \end{pmatrix}\begin{pmatrix}\phi\\ \theta\end{pmatrix}=\varepsilon\begin{pmatrix}\phi\\ \theta\end{pmatrix} \qquad (3)$$

where $S \equiv \sin(\rho)$ and $C \equiv \cos(\rho)$. We consider the case where the even component of the relativistic potential $\mathcal{V}(r) = -De^{-\lambda r}$, where $D$ and $\lambda$ are real parameters. Equation (3) results in the following Schrödinger-like second order differential equation for the upper spinor component

$$\left[-\frac{d^2}{dr^2} + \left(\frac{\alpha D}{T}\right)^2 e^{-2\lambda r} - \frac{\alpha D}{T}\left(\lambda + \frac{2T}{\alpha}\varepsilon\right)e^{-\lambda r} - \frac{\varepsilon^2-1}{\alpha^2}\right]\phi(r) = 0$$

where $T \equiv S/C = \tan(\rho)$. Comparing this with Schrödinger equation for the S-wave nonrelativistic Mörse potential [4] we obtain a correspondence map among the parameters of the two problems. Using the nonrelativistic energy spectrum together with this parameter map, we obtain the following equation for the relativistic spectrum:

$$\varepsilon^2 + \left[\alpha\lambda\left(T\varepsilon/\alpha\lambda - n\right)\right]^2 = 1, \qquad n = 0,1,...,n_{\max} \leq \sqrt{1+T^2}/\alpha\lambda \qquad (4)$$

which can be written as $\cos^2\varphi + \sin^2\varphi = 1$, where $\varepsilon = \cos\varphi$. Therefore, we need to solve the following equation for the angle $\varphi_n$:

$$\frac{T}{\alpha\lambda}\cos\varphi_n - \frac{1}{\alpha\lambda}\sin\varphi_n = n$$

which can be rewritten as:

$$\sin(\rho)\cos(\varphi_n) - \cos(\rho)\sin(\varphi_n) = n\alpha\lambda\cos(\rho)$$

Its solution is $\varphi_n = \rho - \sin^{-1}\left[n\alpha\lambda\cos(\rho)\right]$ resulting in the following elegant analytic formula for the energy spectrum:

$$\varepsilon_n = \cos\left\{\rho - \sin^{-1}\left[n\alpha\lambda\cos(\rho)\right]\right\} \tag{5}$$

This is an alternative but equivalent result to that obtained in reference [2] which reads

$$\varepsilon_n = \frac{1}{1+\tan(\rho)^2}\left[n\alpha\lambda\tan(\rho) + \sqrt{1+\tan(\rho)^2 - (n\alpha\lambda)^2}\right] \tag{6}$$

The spectrum in (6) is a direct solution of equation (4) which is quadratic in $\varepsilon$. The new formula (5) gives a more transparent picture of the overall property of the spectrum, its bounds, and general behavior with the bound state index $n$.